\documentclass[10pt,twocolumn]{IEEEtran}

\usepackage{cite}
\usepackage{graphicx}
\usepackage{epsfig}
\usepackage{psfrag}
\usepackage{subfigure}
\usepackage{url}
\usepackage{stfloats}
\usepackage{amsmath}
\usepackage{color}
\usepackage{amssymb}
\usepackage{algorithm}
\usepackage{algorithmic}
\usepackage{multirow}
\usepackage{array}
\usepackage{booktabs}
\usepackage{multirow}

\begin{document}
%
\title{TROIKA: A General Framework for Heart Rate Monitoring Using Wrist-Type Photoplethysmographic Signals During Intensive Physical Exercise}

\author{Zhilin Zhang$^*$, \IEEEmembership{Member, IEEE}, Zhouyue Pi, \IEEEmembership{Senior Member, IEEE}, and Benyuan Liu
\thanks{Manuscript received July 10, 2014; revised September 5, 2014; accepted September 9, 2014. \emph{Asterisk indicates corresponding author}.}
\thanks{$^*$Z. Zhang and Z. Pi are with the Emerging Technology Lab, Samsung Research America -- Dallas, 1301 East Lookout Drive, Richardson, TX 75082, USA. Email: zhilinzhang@ieee.org (Z. Zhang). }
\thanks{B. Liu is with the Science and Technology on Automatic Target Recognition Laboratory, National University of Defense Technology, Changsha, Hunan 410074, P.R. China. }
}

\markboth{Published in IEEE Transactions on Biomedical Engineering, Vol. 62, No. 2, pp. 522-531, February 2015}{Zhang \MakeLowercase{\textit{et al.}}: }

\maketitle

\begin{abstract}

Heart rate monitoring using wrist-type photoplethysmographic (PPG) signals during subjects' intensive exercise is a difficult problem, since the signals are contaminated by extremely strong motion artifacts caused by subjects' hand movements. So far few works have studied this problem. In this work, a general framework, termed TROIKA, is proposed, which consists of signal decomposiTion for denoising, sparse signal RecOnstructIon for high-resolution spectrum estimation, and spectral peaK trAcking with verification.
The TROIKA framework has high estimation accuracy and is robust to strong motion artifacts. Many variants can be straightforwardly derived from this framework. Experimental results on datasets recorded from 12 subjects during fast running at the peak speed of 15 km/hour showed that the average absolute error of heart rate estimation was 2.34 beat per minute (BPM), and the Pearson correlation between the estimates and the ground-truth of heart rate was 0.992. This framework is of great values to wearable devices such as smart-watches which use PPG signals to monitor heart rate for fitness.
\end{abstract}


%

\IEEEpeerreviewmaketitle

\section{Introduction}

Heart rate (HR) monitoring for fitness is important for exercisers to control their training load. Thus it is a key feature in many wearable devices, such as Samsung Gear Fit, Atlas Fitness Tracker, and Mio Alpha Heart Rate Sport Watch. These devices estimate HR in real time using photoplethysmographic (PPG) signals recorded from wearers' wrist.

PPG signals \cite{allen2007PPG,kamal1989skin} are optically obtained by pulse oximeters. Embedded in a wearable device, a pulse oximeter illuminates a wearer's skin using a  light-emitting diode (LED), and measures intensity changes in the light reflected from skin, forming a PPG signal. The periodicity of the PPG signal corresponds to the cardiac rhythm, and thus HR can be estimated using the PPG signal.

However, PPG signals are vulnerable to motion artifacts (MA), which strongly interfere with HR monitoring in fitness. Many signal processing techniques have been proposed to remove MA.

One technique is independent component analysis (ICA). For example, Kim \emph{et al}. \cite{kim2006ICA} suggested to use a basic ICA algorithm and block interleaving to remove MA.  Later, Krishnan \emph{et al}. \cite{krishnan2010TwoStageICA} proposed using frequency-domain based ICA. However, the key assumption in ICA, namely statistical independence or uncorrelation, does not hold in PPG signals contaminated by MA \cite{yao2005ICAstudy}. Thus MA removal is not satisfactory. Besides, using ICA needs multiple PPG sensors, which may not be feasible in some wearable devices.

Another technique is adaptive noise cancelation (ANC) \cite{ram2012AS_LMS,yousefi2013ANC}. For example, Ram \emph{et al}. \cite{ram2012AS_LMS} proposed using ANC to remove MA, where the reference signal was constructed from fast fourier transform (FFT), singular value decomposition, or ICA of artifact-contaminated PPG signals. However, one limitation in this technique is that the artifact-removal performance of ANC is sensitive to the reference signal, and reconstructing qualified reference signals is difficult when subjects are exercising.

Acceleration data are also  helpful to remove MA. For example, Fukushima \emph{et al}. \cite{fukushima2012SpectrumSubtraction} suggested a spectrum subtraction technique to remove the spectrum of acceleration data from that of a PPG signal. Acceleration data can be also used to reconstruct  the observation model for Kalman filtering \cite{lee2010kalman} to remove MA. However, acceleration data (in three-axis) reflect hand movement in 3-D space, while MA in a PPG signal originates from changes of the gap between skin and a pulse oximeter's surface. Thus, in the presence of irregular and complicated hand movements, merely exploiting acceleration data may not gain large benefits.

Other MA-removal techniques include electronic processing methodology \cite{hayes2001new}, time-frequency analysis \cite{yan2005WVdistribution}, wavelet denoising \cite{raghuram2010wavelet}, and empirical mode decomposition \cite{sun2012robust}, to name a few.

Most of these techniques were proposed for clinical scenarios, when wearers performed small motions, such as finger movements \cite{reddy2009Fourier,ram2012AS_LMS,yousefi2013ANC,krishnan2010TwoStageICA} and walking \cite{yousefi2013ANC}. In these scenarios MA was not strong. Thus these  techniques may not work well when wearers perform intensive physical exercise.

Only a few techniques were proposed for HR monitoring in fitness. But most of them focused on slowly running (speed slower than 8 km/hour) \cite{poh2010motion,yousefi2013ANC}, and PPG data were recorded from fingertip \cite{lopez2012heuristic,yousefi2013ANC} or ear \cite{poh2010motion}. In these cases, MA was not very strong and even exhibited characteristics facilitating HR estimation \cite{lopez2012heuristic}.

Also, few works studied MA removal in PPG signals recorded from wrist. Comparing to fingertip and earlobe, wrist can cause much stronger and complicated MA due to large flexibility of wrist and loose interface between pulse oximeter and skin. However, recording PPG from wrist greatly facilitates design of wearable devices and maximizes user experience. Furthermore, estimating HR from wrist-type PPG signals becomes a popular feature in smart-watch type devices. Thus developing high performance HR monitoring algorithms for wrist-type PPG signals is of great values.

This work focuses on HR monitoring using wrist-type PPG signals when wearers perform intensive physical activities. A
novel framework is proposed. Termed TROIKA, it consists of three key parts, namely signal decomposiTion, sparse signal RecOnstructIon, and spectral peaK trAcking. Signal decomposition aims to partially remove MA in a raw PPG signal and sparsify its spectrum. Sparse signal reconstruction aims to calculate a high-resolution spectrum of the PPG signal, which is robust to noise interference and is advantageous over traditional non-parametric spectrum estimation algorithms and model-based line spectrum estimation algorithms. Spectral peak tracking is also a necessary part of the TROIKA framework, which seeks the spectral peak corresponding to HR. To further overcome strong interference from MA and complement the peak selection approach, some decision mechanisms are designed to verify the selected spectral peak.

Experimental results on datasets recorded from 12 subjects during fast running at the peak speed of 15 km/hour showed that the proposed framework has high estimation performance with the average absolute error being only 2.34 beat per minute (BPM).

The rest of the paper is organized as follows. Section \ref{sec:motivation} states the HR monitoring problem and our motivations. Section \ref{sec:TROIKA} presents the TROIKA framework. Section \ref{sec:experiments} describes recorded datasets, experiment settings, and experimental results. Conclusion is given in the last section.

Matlab codes and datasets used in experiments are available at the first author's homepage: \url{https://sites.google.com/site/researchbyzhang/}. A preliminary work has been published in \cite{Zhang2014GlobalSIP}.

\section{Problem Statement and Motivations}
\label{sec:motivation}

When MA is strong, estimating HR in the time domain is difficult. A natural approach is to estimate HR using power spectrum estimation. One widely used algorithm is the Periodogram algorithm computed via Fast Fourier Transform (FFT) \cite{stoica2005spectral}. However, Periodogram has some drawbacks.

First, Periodogram is an inconsistent spectrum estimator, and has high variance \cite{stoica2005spectral}. Second, Periodogram has serious leakage effect \cite{stoica2005spectral}. Thus a dominant spectral peak can lead to an estimated spectrum that contains power in frequency bands where there should be no power. The effect can make a weak spectral peak associated with HR smeared, if a dominant spectral peak associated with MA is nearby.

\begin{figure}[t]
\centering
\includegraphics[width=9cm,height=7cm]{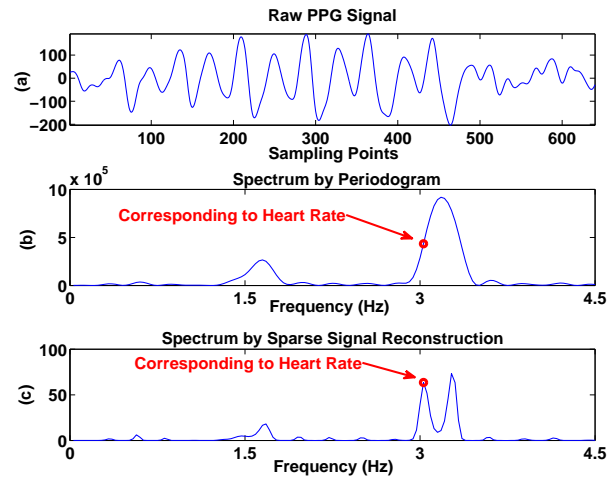}
\caption{An example showing the leakage effect of Periodogram. (a) A raw PPG signal sampled at 125 Hz and band-pass filtered from 0.4 Hz to 5 Hz. It has a dominant quasi-periodic signal component caused by rhythmic hand  swing, and a weak signal component caused by heartbeat. The two components have close periods. (b) The PPG's spectrum calculated by Periodogram. The spectral peak corresponding to heart rate (indicated by a circle) cannot be distinguished from the one corresponding to the hand swing. (c) The PPG's spectrum calculated by FOCUSS \cite{Gorodnitsky1997}, an SSR algorithm. The two spectral peaks are clearly separated. The ground-truth heart rate is obtained from simultaneously recorded ECG. }
\label{fig:FFTvsCS}
\end{figure}

Figure~\ref{fig:FFTvsCS}(a) and (b) shows a raw PPG signal and its spectrum calculated by Periodogram, respectively. The signal has a dominant quasi-periodic signal component resulting from rhythmic hand swing, and another signal component corresponding to heartbeat. The two components have close periods. Due to the leakage effect, the spectral peak associated with the HR cannot be separated from the  peak associated with the hand swing rhythm. Thus an error in HR estimation could occur.

This example motivates using high-resolution line spectrum estimation. However, conventional high-resolution line spectrum estimation algorithms such as MUSIC require model order selection, which is difficult for MA-contaminated PPG signals, since the spectra of MA are complicated and time-varying.

The limitations of  conventional non-parametric spectrum estimation and high-resolution line spectrum estimation motivate us to   use   sparse signal reconstruction (SSR) \cite{Gorodnitsky1997,Donoho2006compressed,Eladbook}. Recently it is found that SSR-based spectrum estimation has a number of advantages over traditional spectrum estimation \cite{duarte2013spectral}. Compared to  nonparametric spectrum estimation methods such as Periodogram,  the SSR-based spectrum estimation features high spectrum resolution, low estimation variance, and increased robustness; compared to conventional line spectral estimation methods, the SSR-based spectrum estimation does not require model selection and has improved estimation performance. Figure~\ref{fig:FFTvsCS} (c) shows the spectrum calculated by an SSR algorithm, where the spectral peak associated with the HR is clearly separated from the spectral peak associated with the hand swing.

Note that SSR requires that the spectrum to estimate is sparse or compressive; that is, the spectrum has a few spectrum coefficients of large nonzero values, while other coefficients have zero values or nearly zero values.
Since MA-contaminated PPG signals may not have sparse/compressive spectra, SSR needs preprocessing to sparsify the spectra.

\begin{figure}[t]
\centering
\includegraphics[width=9cm,height=7cm]{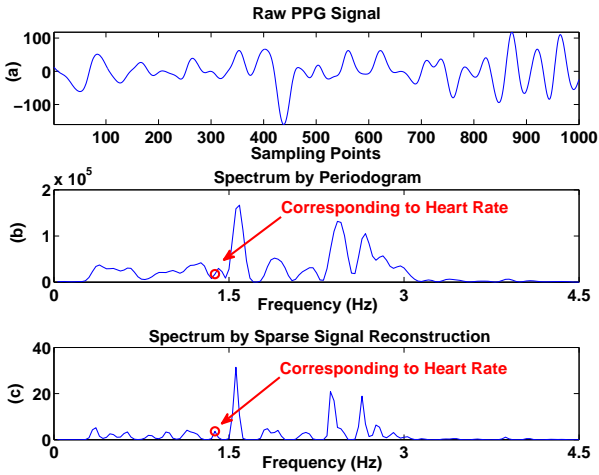}
\caption{An example showing many spectral peaks exhibit when MA occurs. (a) A PPG signal after bandpass filtered from 0.4 Hz to 5 Hz. (b) The spectrum calculated by Periodogram. (c) The spectrum calculated by FOCUSS. The circles in (b) and (c) indicate the frequency location corresponding to the true heart rate (which is calculated from simultaneously recorded ECG).}
\label{fig:manyPeak}
\end{figure}

\begin{figure}[t]
\centering
\includegraphics[width=9cm,height=7cm]{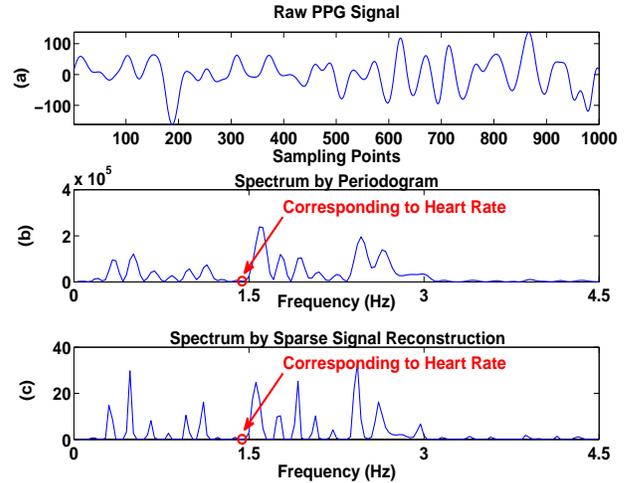}
\caption{An example showing a case when the spectral peak associated with the heart rate does not exhibit. (a) A raw PPG signal after bandpass filtered from 0.4 Hz to 5 Hz. (b) The spectrum calculated by Periodogram. (c) The spectrum calculated by FOCUSS. The circles in (b) and (c) indicate the frequency location corresponding to the true heart rate (which is calculated from simultaneously recorded ECG).}
\label{fig:noHRpeak}
\end{figure}

However, a high-resolution spectrum estimation algorithm is insufficient to solve the problem of HR monitoring. A large gap between a pulse oximeter and skin could make the spectral peak associated with HR buried in random spectral fluctuations, as shown in  Figure~\ref{fig:manyPeak}. Finding the spectral peak is difficult. In an extreme situation, a PPG signal may not contain any heartbeat-related signal component, and thus one cannot find the spectral peak associated with HR, as shown in Figure~\ref{fig:noHRpeak}. Therefore some methods are needed to deal with the cases when the corresponding spectral peak is buried in spectral fluctuations or does not exhibit.

Due to the above issues,  a satisfactory framework for HR monitoring during intensive exercise should consists of three parts: denoising, high-resolution spectrum estimation, and spectral peak tracking (including peak selection and verification). Thus we propose the TROIKA framework. Details of the framework are given in the next section.

\begin{figure}[t]
\centering
\includegraphics[width=9cm,height=4.5cm]{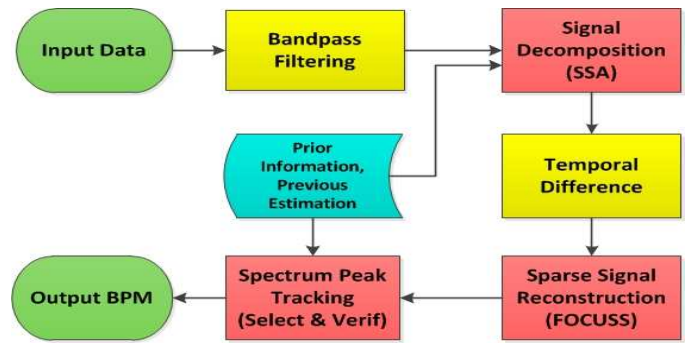}
\caption{Flowchart of the TROIKA framework. Signal decomposition, sparse signal reconstruction, and spectral peak tracking are the three key parts.}
\label{fig:framework}
\end{figure}

\section{The TROIKA Framework}
\label{sec:TROIKA}

The TROIKA framework consists of three key parts: signal decomposition, SSR, and spectral peak tracking. Signal decomposition is used for denoising and sparsifying spectra of PPG signals. SSR yields high-resolution spectrum estimation. The spectral peak tracking part selects correct spectral peaks and deal with the non-existence cases of the peaks. The flowchart of the framework (including other auxiliary signal processing parts) is shown in Figure~\ref{fig:framework}.

TROIKA is used for a single-channel PPG signal and simultaneously recorded acceleration data. A time window of $T$ seconds is sliding on the  signals with incremental step of $S$ seconds (generally $S \leq T/2$). The framework estimates HR in each time window. In our experiments $T=8$ and $S=2$ \footnote{The exact values of $T$ and $S$ are not important. Choosing relatively large $T$ is to increase spectrum resolution, while choosing small $S$ is to make the calculated heart rates in two successive time windows close to each other.}.

Before TROIKA starts, the raw PPG signal and the acceleration signal in a given time window are first bandpass filtered from 0.4 Hz and 5 Hz. This preprocessing removes noise and MA outside of the frequency band of interest, which also sparsifies the spectra.

\subsection{Signal Decomposition}

Signal decomposition is an effective denoising methodology. Generally speaking, it decomposes a signal $\mathbf{y}$ into a number of components as follows
\begin{eqnarray}
\mathbf{y} = \sum_{i=1}^Q \mathbf{y}_i.
\end{eqnarray}
Then, based on some criteria, the components associated with noise/interference are recognized and removed from the above expression. The reconstructed signal is thus noise- and interference-free.

There are many signal decomposition approaches, such as singular spectrum analysis (SSA) \cite{golyandina2001SSA,harris2010filtering}, single-channel independent component analysis (SCICA) \cite{james2003extracting,davies2007source},  and empirical mode decomposition (EMD) \cite{huang1998empirical}. These approaches decompose signals based on different criteria and procedures.

For illustration, we chose SSA in our experiments. SSA decomposes a time series into oscillatory components and noise. It includes four steps: Embedding, Singular Value Decomposition (SVD), Grouping, and Reconstruction.

In the Embedding Step, a time series $\mathbf{y} \triangleq [y_1,\cdots,y_M]^T$ is mapped into an $L \times K$ matrix ($K=M-L+1, L<M/2$), called L-trajectory matrix \footnote{The parameter $L$ is defined by users. It is suggested to set $L$ close to $M/2$ for stable decomposition \cite{golyandina2001SSA}.},
\begin{eqnarray}
\mathbf{Y}  & \triangleq & \left[ \begin{array}{cccc}
y_1      &      y_2       &      \cdots     &  y_K     \\
y_2      &      y_3       &       \cdots    &  y_{K+1} \\
\vdots   &      \vdots    &      \ddots     &  \vdots  \\
y_L      &      y_{L+1}   &    \ldots       &  y_M   \end{array} \right].
\end{eqnarray}

In the SVD Step, the L-trajectory matrix is decomposed by SVD as follows,
\begin{eqnarray}
\mathbf{Y} = \sum_{i=1}^d \mathbf{Y}_i, \quad d \triangleq  \min\{L,K\}
\end{eqnarray}
where $\mathbf{Y}_i = \sigma_i \mathbf{u}_i \mathbf{v}_i^T$, and $\sigma_i, \mathbf{u}_i, \mathbf{v}_i$ are the $i$th singular value, the corresponding left-singular vector and the corresponding right-singular vector, respectively.

In the Grouping Step, the $d$ rank-one matrix $\mathbf{Y}_i$ are assigned into $g$ groups. That is, the set of indices $\{1,\cdots,d\}$ is partitioned into $g$ disjoint subsets $\{I_1,\cdots,I_g\}$ ($g \leq d$) and
\begin{eqnarray}
\mathbf{Y} = \sum_{p=1}^g \mathbf{Y}_{I_p},
\end{eqnarray}
with $\mathbf{Y}_{I_p} = \sum_{t \in I_p} \mathbf{Y}_t$. The rank-one matrices in each group $\mathbf{Y}_{I_p}$ satisfy some common characteristics (such as their corresponding  oscillatory components after reconstruction have the same frequency or exhibit harmonic relation).

In the Reconstruction Step, each group $\mathbf{Y}_{I_p}$ is used to reconstruct a time series $\widetilde{\mathbf{y}}_p$ with the length $M$ by a so-called diagonal averaging procedure \cite{golyandina2001SSA}. Thus the original signal $\mathbf{y}$ can be re-expressed as a sum of $g$ time series, i.e.
\begin{eqnarray}
\mathbf{y} = \sum_{p=1}^g \widetilde{\mathbf{y}}_p.
\end{eqnarray}
Once having identified noise time series $\{ \widetilde{\mathbf{y}}_p, p\in \mathcal{I}_{\mathrm{noise}} \}$, where $\mathcal{I}_{\mathrm{noise}}$ indicates the set of indices of all identified noise time series, one can reconstruct a noise-free signal as follows,
\begin{eqnarray}
\mathbf{y}_{\mathrm{recon}} = \sum_{p \not \in \mathcal{I}_{\mathrm{noise}}} \widetilde{\mathbf{y}}_p.
\end{eqnarray}

In our framework, SSA partially removes MA frequency components from 0.4 Hz to 5 Hz. (MA components with frequencies outside of this range have already been removed by a bandpass filter.) This also sparsifies the spectrum coefficients in the range from 0.4 Hz to 5 Hz, alleviating the difficulty of SSR. To recognize the MA components in the SSA decomposition, we exploit information from acceleration data. Details are given below.

First, in current time window Periodogram \footnote{Here using Periodogram is to save computational load. Advanced spectrum estimation algorithms are not necessarily needed, since the goal is to select dominant frequencies of MA in acceleration data.} is used to calculate the spectrum of each channel of acceleration data \footnote{Three-axis acceleration data were used in our experiments.}, from which dominant frequencies are determined. The dominant frequencies are the ones corresponding to the spectral peaks with amplitude larger than 50\% of the maximum amplitude in a given spectrum. Denote by $\mathcal{F}_{\mathrm{acc}}$ the set of location indexes of selected dominant frequencies in spectra.

Then SSA is performed on the PPG signal in current time window. After SVD, the grouping is automatically finished by clustering singular values as described in \cite[pp. 66]{golyandina2001SSA}. Thus the original PPG signal is decomposed into a number of time series. Next, we remove the time series whose  dominant frequencies have  location indexes in $\mathcal{F}_{\mathrm{acc}}$. The remained time series are used to reconstruct a cleansed PPG signal.

Note that some selected dominant frequencies  may be close to the fundamental and harmonic frequencies of heartbeat. Thus the above procedure may remove the signal components associated with the heartbeat. Some modifications to the above procedure are needed. Denote by $N_{\mathrm{p}}$ the location indexes of fundamental and harmonic frequencies of the heartbeat estimated in the previous time window. We exclude $\{ N_{\mathrm{p}}-\Delta, \cdots, N_{\mathrm{p}}-1, N_{\mathrm{p}}, N_{\mathrm{p}}+1, \cdots, N_{\mathrm{p}}+\Delta    \}$ from $\mathcal{F}_{\mathrm{acc}}$, forming a refined set of location indexes of dominant frequencies, denoted by  $\widetilde{\mathcal{F}}_{\mathrm{acc}}$. $\Delta$ is a small positive integer. Then the above SSA procedure is performed using $\widetilde{\mathcal{F}}_{\mathrm{acc}}$.

In experiments we set the parameters $L=400$, set the number of FFT points used in Periodogram  to 4096, and set $\Delta=10$. Since the sampling rate was 125 Hz in our experiments, with 4096 FFT points, the parameter value $\Delta=10$ corresponds to about 0.3 Hz.

\subsection{Temporal Difference Operation}

To further improve robustness of the framework, we suggest temporally differentiating the cleansed PPG signal output by SSA and then performing SSR-based spectrum estimation.

For a periodic time series $h = [h(1),h(2),\cdots,h(M)]$, its first-order difference, defined as $h'\triangleq [h(2)-h(1), h(3)-h(2), \cdots, h(M)-h(M-1)]$, maintains the fundamental frequency and harmonic frequencies. The second-order difference of $h$, i.e. the first-order difference of $h'$, also maintains the fundamental frequency and the harmonic frequencies. As long as $k$ is not large, the spectrum of the $k$th-order difference of the time series  significantly presents the fundamental frequency and its harmonic frequencies. In contrast, this phenomenon is not observed from a non-periodic time series.

Note that the PPG component associated with HR is approximately periodic in a short time window, while MA is generally aperiodic (except to the MA components corresponding to rhythmic hand swing). Therefore, the temporal difference operation can make the heartbeat fundamental and harmonic spectral peaks more prominent, while suppressing random spectrum fluctuations.

In experiments we calculated the second-order difference of the cleansed PPG signal output by SSA.

\subsection{SSR}

SSR \cite{Gorodnitsky1997,cotter2005sparse,Donoho2006compressed,Eladbook} is an emerging signal processing technique, showing great potentials in many application fields. The basic SSR model can be expressed as follows,
\begin{eqnarray}
\mathbf{y} = \mathbf{\Phi} \mathbf{x} + \mathbf{v}.
\label{eq:basicModel}
\end{eqnarray}
Here $\mathbf{\Phi}$ is a known basis matrix of the size $M \times N$. The vector $\mathbf{y}$ is an observed signal of the size $M \times 1$.  $\mathbf{x}$ is an unknown solution vector which is assumed to be sparse or compressive, i.e. most elements of $\mathbf{x}$  are zero or nearly zero, while only a few elements have large nonzero values. $\mathbf{v}$ is an unknown noise vector. The goal is to find the sparsest vector $\mathbf{x}$ based on $\mathbf{y}$ and $\mathbf{\Phi}$. This can be done by solving the following optimization problem
\begin{eqnarray}
\widehat{\mathbf{x}} \leftarrow \min_{\mathbf{x}} \|\mathbf{y} - \mathbf{\Phi} \mathbf{x}\|_2^2 + \lambda g(\mathbf{x})
\label{eq:basicSolution}
\end{eqnarray}
where $\lambda$ is a regularization parameter, and $g(\mathbf{x})$ is a penalty function encouraging the sparsity of the solution $\widehat{\mathbf{x}}$. Some popular penalty functions are the $\ell_1$-norm function, i.e. $\| \mathbf{x} \|_1$ \cite{lasso}, or the $\ell_p$-norm function ($0<p<1$), i.e. $\| \mathbf{x} \|_p$ \cite{Gorodnitsky1997}.

It is interesting to see that SSR can be used in spectrum estimation \cite{Gorodnitsky1997,duarte2013spectral}. Let $\mathbf{y} \in \mathbb{R}^{M \times 1}$ be a real-valued signal. Constructing the matrix $\mathbf{\Phi} \in \mathbb{C}^{M \times N}$ such that its $(m,n)$-th element ${\Phi}_{m,n}$ is given by
\begin{eqnarray}
\Phi_{m,n} = e^{j \frac{2\pi}{N}mn}, \; m = 0,\cdots,M-1; \, n=0,\cdots,N-1
\label{eq:basis}
\end{eqnarray}
the solution (\ref{eq:basicSolution}) leads to a sparse spectrum of the signal $\mathbf{y}$. Specifically, denoting the spectrum by $\mathbf{s} \in \mathbb{R}^{N\times 1}$, its $k$-th element, $s_k$, is given by
\begin{eqnarray}
s_k = | \widehat{x}_k |^2, \quad k=1,\cdots,N
\end{eqnarray}
where $\widehat{x}_k$ is the $k$-th element of $\widehat{\mathbf{x}} \in \mathbb{C}^{N \times 1}$.

When performing SSR with the model (\ref{eq:basicModel})-(\ref{eq:basis}), the size of $\mathbf{\Phi}$ is very large, implying large computational load. However, a simple pruning procedure to $\mathbf{\Phi}$ before running SSR algorithms can largely reduce computational load. Let $f_s$ denote the sampling rate. Equation (\ref{eq:basis}) indicates that the whole spectrum $[0, f_s]$ is divided into $N$ equal frequency bins. A frequency bin with location index $N_f \in \{1,2,\cdots,N\}$ corresponds to a physical frequency
\begin{eqnarray}
f = \frac{N_f-1}{N} f_s \quad (\mathrm{Hz})
\label{eq:physicalF}
\end{eqnarray}
For example, the first frequency bin with location index $N_f=1$ corresponds to 0 Hz.

In our experiments the signal has been bandpass filtered from 0.4 Hz to 5 Hz. Thus only the spectrum coefficients with the location indexes $N_f$ satisfying $N_f \in \mathcal{I}_{\Phi} $ have nonzero values, where $\mathcal{I}_{\Phi}$ is given by
\begin{eqnarray}
\mathcal{I}_{\Phi} &\triangleq & \Big[\frac{0.4}{f_s}N + 1 - \Delta_{f1}, \; \frac{5}{f_s}N+1+\Delta_{f2} \Big]  \nonumber \\
&  \cup &  \Big[N-\frac{5}{f_s}N+1-\Delta_{f2}, \; N-\frac{0.4}{f_s}N+1+\Delta_{f1} \Big] \nonumber \\
\end{eqnarray}
Here using $\Delta_{f1}$ and $\Delta_{f2}$ is to consider the width of the transition band of a practical bandpass filter. Thus the columns of $\Phi$ with indexes $N_f \not\in \mathcal{I}_{\Phi}$ can be pruned out before performing SSR. This can largely reduce computational load. In experiments we set $\Delta_{f1} = \frac{0.4}{f_s}N-1$ and $\Delta_{f2}=\frac{2}{f_s}N$.

Although the proposed framework is general, not restricted to a specific algorithm, carefully choosing an SSR algorithm is helpful to improve the performance of HR monitoring. Note that when $M \ll N$, the columns of the basis matrix $\mathbf{\Phi}$ are highly correlated \cite{duarte2013spectral}. In this case some SSR algorithms do not work well. In experiments we chose the FOCUSS algorithm \cite{Gorodnitsky1997} for illustrations. The algorithm is robust when columns of $\mathbf{\Phi}$ are highly correlated, and has been used widely in source localization and direction-of-arrival estimation. Details on the parameter settings are given in Section \ref{sec:experiments}.

\textbf{Remark 1}: As stated in Section \ref{sec:motivation}, the SSR-based spectrum estimation has a number of advantages over traditional  spectrum estimation algorithms \cite{Gorodnitsky1997,duarte2013spectral}. However, a key assumption of SSR is that $\mathbf{x}$ should be sparse or compressive. Otherwise, the performance is degraded. When complicated intensive hand movements occur, the spectrum of a PPG signal is non-sparse, causing poor performance of SSR. Thus the preprocessing procedures including bandpass filtering, signal decomposition, and temporal difference operation are important, since they can sparsify spectrum coefficients, alleviating the difficulty in SSR.

\textbf{Remark 2}: Since an input PPG signal is first bandpass-filtered from 0.4 Hz to 5 Hz in the framework, one may think the number of nonzero spectrum coefficients is small enough and the PPG spectrum is  sparse. But it should be noted that when columns of $\mathbf{\Phi}$ are highly correlated (as in our problem), even a small number of nonzero spectrum coefficients can bring difficulty to SSR. Hence it is preferred to further remove the significantly nonzero spectrum coefficients associated with MA after bandpass filtering.

\textbf{Remark 3}: When using SSR to estimate power spectrum of a signal, setting $N$ requires some considerations. Increasing $N$ can reduce the off-grid effect in spectrum estimation \footnote{The off-grid effect refers to the case when a frequency of a signal does not locate on any of the frequency grids $\frac{n}{N}f_s, n=0,\cdots,N-1$.} and thus can potentially increase HR estimation accuracy. However, large $N$ increases the correlation among columns of $\mathbf{\Phi}$, thus potentially decreasing SSR performance. Therefore we suggest  setting $N$ such that each frequency grid $\frac{f_s}{N}60$ corresponds to about 1 BPM. Since in our experiments $f_s=125$ Hz, we set $N=4096$. Also, since each time window last $T=8$ seconds, $M=8\times 125=1000$.

\subsection{Spectral Peak Tracking}

Spectral peak tracking is another key part in TROIKA. It exploits the frequency harmonic relation of HR, and the observation that HR values in two successive time windows are very close if the two time windows overlap largely. In fact,  our experiments showed that in many cases the spectral peak associated with HR keeps its location unchanged in two successive time windows.

The spectral peak tracking consists of initialization, peak selection, and verification.

\subsubsection{Initialization}

In the initialization stage wearers are required to reduce hand motions as much as possible for several seconds (2 or 3 seconds are enough). Thus HR can be estimated by choosing the highest spectral peak in a PPG spectrum during this stage. Without this stage, there may be no way to estimate HR, since a PPG spectrum contains many peaks and there is no prior information to help find the correct one (see Fig.~\ref{fig:manyPeak}).

\subsubsection{Peak Selection}
\label{subsubsec:peakSelection}

Denote by $N_{\mathrm{prev}}$ the frequency location index of HR estimated in the previous time window. We set a search range for the fundamental frequency of HR in current time window, denoted by $R_0 = [ N_{\mathrm{prev}}-\Delta_s, \cdots, N_{\mathrm{prev}}+\Delta_s]$, where $\Delta_s$ is a small positive integer. In our experiments we set $\Delta_s = 16$. Also, we set another search range for the first-order harmonic frequency of HR, denoted by $R_1 = [ 2(N_{\mathrm{prev}}-\Delta_s-1)+1, \cdots,   2(N_{\mathrm{prev}}+\Delta_s-1)+1 ]$ \footnote{Remind that the first frequency bin in a spectrum corresponds to 0 Hz.}. In each search range, we select no more than three highest peaks with amplitude no less than a threshold $\eta$. In our experiments $\eta$ was set to 30\% of the highest peak amplitude in $R_0$. This setting is helpful to remove spectral fluctuations. Denote the frequency location indexes of the three peaks in $R_0$ by $N_i^0 (i=1,2,3)$, and the indexes of the three peaks in $R_1$ by $N_j^1 (j=1,2,3)$.

Let $\widehat{N}$ be the location index of the selected spectral peak. We consider three cases.
\begin{description}
  \item[\textbf{Case 1} ] ~If there exists a peak-pair $(N_i^0, N_j^1) (i,j \in\{1,2,3\})$ with a harmonic relation, then $N_i^0$ is the  frequency location index of HR, and thus
\begin{eqnarray}
\widehat{N} \leftarrow  N_i^0
\end{eqnarray}

  \item[\textbf{Case 2} ] ~If there is no such a peak-pair (due to MA), based on the observation stated previously, we select $\widehat{N}$ as follows
\begin{eqnarray}
\widehat{N} \leftarrow \arg\min_{N_f} \{ | N_f - N_{\mathrm{prev}}| \}
\end{eqnarray}
where $N_f \in \{ N_1^0, N_2^0, N_3^0, \frac{N_1^1-1}{2}+1,  \frac{N_2^1-1}{2}+1, \frac{N_3^1-1}{2}+1   \}$ \footnote{Because the first frequency bin corresponds to 0 Hz, the calculation of $(N_1^1-1)/2+1$ is used, instead of $N_1^1/2$. Similarly we have $(N_2^1-1)/2+1$ and $(N_3^1-1)/2+1$.}.

  \item[\textbf{Case 3} ] ~If no peaks are found in either search range, we have
\begin{eqnarray}
\widehat{N} \leftarrow  N_{\mathrm{prev}}
\label{eq:case3}
\end{eqnarray}
  which is based on the observation that the spectral peak associated with HR often keeps the same location in two successive time windows.
\end{description}

\subsubsection{Verification}

The peak selection method sometimes can wrongly track the spectral peak associated with MA or spectral fluctuations. Thus a verification stage is necessary. Several rules are designed to prevent the wrong tracking.

The first  rule is to prevent a large change in the estimated BPM values in two successive time windows. The change of BPM values in two successive time windows rarely exceeds 10 BPM. Therefore, once the change exceeds 10 BPM, a regularization is performed as follows
\begin{eqnarray}
N_{\mathrm{cur}} =
\left\{
\begin{array}{ll}
N_{\mathrm{prev}} + \tau  & \quad \textrm{if $\widehat{N} - N_{\mathrm{prev}} \geq \theta$}   \\
N_{\mathrm{prev}} - \tau  & \quad \textrm{if $\widehat{N} - N_{\mathrm{prev}} \leq - \theta$}   \\
\widehat{N}  & \quad \textrm{otherwise}   \\
\end{array} \right.
\end{eqnarray}
where $\widehat{N}$ is the selected spectral peak location in Section \ref{subsubsec:peakSelection}, and $N_{\mathrm{cur}}$ is the regularized spectral peak location. Since in our experiments the whole spectrum $[0, 125]$ (Hz) is divided into $N=4096$ grids, a change of 6 locations in the spectrum means a change of about 11 BPM. Thus we set $\theta = 6$. The parameter $\tau$ can be set to any small positive integer, such as $\tau = 2$.

The second  rule is to prevent from losing tracking over long time. If strong MA interference continues over long time, the spectral peak selection by (\ref{eq:case3}) could happen in many successive time windows. Thus the spectral peak  associated with  HR is  lost and may not be found back. To avoid this situation, if $\widehat{N} = N_{\mathrm{prev}}$ for $h$ successive time windows, then
\begin{eqnarray}
N_{\mathrm{cur}} = N_{\mathrm{prev}} + 2 \cdot N_{\mathrm{Trend}}
\end{eqnarray}
where $N_{\mathrm{Trend}} \in \{-1, 0, 1\}$ indicating the change direction of the HR peak location. Also, when this situation occurs,  the search range is broadened by setting $\Delta_s = 20$. (If the situation does not occur, $\Delta_s$ is set to the original value, i.e. $\Delta_s=16$, since narrower search range is helpful to avoid interference from non-HR spectral peaks.) In  experiments we set $h=3$, and calculated $N_{\mathrm{Trend}}$ as follows
\begin{eqnarray}
N_{\mathrm{Trend}} =
\left\{
\begin{array}{ll}
1  & \quad \textrm{if $\mathrm{BPM}_{\mathrm{predict}} - \mathrm{BPM}_{\mathrm{prev}} \geq 3$}   \\
-1  & \quad \textrm{if $\mathrm{BPM}_{\mathrm{predict}} - \mathrm{BPM}_{\mathrm{prev}} \leq -3$}   \\
0  & \quad \mathrm{otherwise}
\end{array} \right.
\end{eqnarray}
where $\mathrm{BPM}_{\mathrm{prev}}$ was the estimated BPM in the previous time window, and $\mathrm{BPM}_{\mathrm{predict}}$ was the predicted BPM of the current time window. The predicted BPM was obtained by using the third-order Polynomial curve fitting on the estimated BPM values in previous 20 time windows.

\subsection{Remarks}

TROIKA is a general framework. Although we adopt specific algorithms for signal decomposition and SSR, other signal decomposition algorithms and SSR algorithms can be used as well. For example, sparse Bayesian learning \cite{Zhang2013TSP} is an alternative to FOCUSS, and EMD is an alternative to SSA.

TROIKA has some user-defined parameters. In previous sections we assigned values to these parameters by heuristics. Experiments showed that TROIKA's performance is robust to these values; other  values can yield similar performance (see the following section). However, we emphasize that these parameter values are assigned based on the sampling rate used in our experiments ($f_s=125$ Hz). When the sampling rate changes much, several parameters' values should be adjusted proportionally. For example, when $f_s$ is 25 Hz instead of 125 Hz, the number of frequency bins, $N$, should take about $1/5$ of the original value such that other parameters' values (e.g. $\Delta_s$ and $\Delta$) maintain the same physical measures.

\begin{table*}[t]
\renewcommand{\arraystretch}{1.2}
\caption{Average absolute error (Error1) on all 12 subjects' recordings. In the results, the first row was obtained when using the complete TROIKA framework. The second, third, and fourth row was obtained when removing the signal decomposition part, replacing the SSR algorithm with FFT, and removing the spectral peak verification part, respectively. Only the complete TROIKA framework showed robust and high performance.}
\label{table:err1}
\centering
\begin{tabular}{l|c|c|c|c|c|c|c|c|c|c|c|c}
\toprule
       &  Subj 1      & Subj 2   &  Subj 3   &  Subj 4   &Subj 5  &Subj  6  &Subj 7  &Subj  8  &Subj 9  &Subj 10 &Subj 11 & Subj 12\\
\midrule
Err1 (SSA+FOCUSS+Vrf)       &   2.29   &  2.19   &  2.00   &  2.15    & 2.01   &   2.76 & 1.67 & 1.93 & 1.86 & 4.70  & 1.72  & 2.84\\
\hline
Err1 ($\quad\quad\;$FOCUSS+Vrf)  &   4.56    & 4.63     & 3.97     & 2.81     & 2.06    & {\color{red}$\underline{55.2}$}   & 1.84  & 1.75  & 1.84  & 5.86   &  4.92  & 8.76  \\
\hline
Err1 (SSA+FFT$\quad\quad$+Vrf)       &   {\color{red}$\underline{62.73}$}   &  5.55   &  1.99   &  3.25    & 1.11   &  {\color{red}$\underline{55.84}$} & 1.17 & 1.68 & 0.45 & {\color{red}$\underline{12.26}$}  & 1.84  & 2.54 \\
\hline
Err1 (SSA+FOCUSS$\quad\quad$)       &   3.45   &  {\color{red}$\underline{14.00}$}   &  {\color{red}$\underline{24.07}$}   &  2.62    & 2.10   &  {\color{red}$\underline{54.79}$} & 2.97 & 1.77 & 1.92 & {\color{red}$\underline{51.89}$}  & 2.69  & {\color{red}$\underline{60.14}$} \\
\bottomrule
\end{tabular}
\end{table*}

\begin{table*}[t]
\renewcommand{\arraystretch}{1.2}
\caption{Average error percentage (Error2) on all 12 subjects' recordings. Each row of the results corresponds to the same case as in Table~\ref{table:err1}.}
\label{table:err2}
\centering
\begin{tabular}{l|c|c|c|c|c|c|c|c|c|c|c|c}
\toprule
       &  Subj 1      & Subj 2   &  Subj 3   &  Subj 4   &Subj 5  &Subj  6  &Subj 7  &Subj  8  &Subj 9  &Subj 10 &Subj 11 & Subj 12\\
\midrule
Err2  (SSA+FOCUSS+Vrf)       &  1.90\%   &  1.87\%   &  1.66\%  & 1.82\%    &  1.49\%   &  2.25\% & 1.26\% &  1.62\% &  1.59\% &  2.93\% & 1.15\% & 1.99\%\\
\hline
Err2  ($\quad\quad\;$FOCUSS+Vrf)  &  3.22\%    & 3.66\%     & 3.01\%    & 2.39\%     &  1.48\%    & {\color{red}$\underline{39.11\%}$}   & 1.39\%  & 1.45\%   & 1.53\%   & 3.59\%   & 3.28\%   &  6.08\% \\
\hline
Err2  (SSA+FFT$\quad\quad$+Vrf)       &  {\color{red}$\underline{42.79\%}$}   &  5.13\%   &  1.80\%  & 2.93\%    &  0.88\%   &  {\color{red}$\underline{39.50\%}$} & 0.89\% &  1.57\% &  0.37\% &  7.44\% & 1.23\% & 1.73\%\\
\hline
Err2  (SSA+FOCUSS$\quad\quad$)       &  2.69\%   &  {\color{red}$\underline{12.91\%}$}   &  {\color{red}$\underline{19.36\%}$}  & 2.30\%    &  1.50\%   &  {\color{red}$\underline{38.87\%}$} & 2.14\% &  1.48\% &  1.53\% &  {\color{red}$\underline{32.82\%}$} & 1.79\% & {\color{red}$\underline{41.00\%}$} \\
\bottomrule
\end{tabular}
\end{table*}

\section{Experimental Results}
\label{sec:experiments}

\subsection{Data Recording}

We simultaneously recorded a single-channel PPG signal, a three-axis acceleration signal, and an ECG signal from 12 male subjects with yellow skin and ages ranging from 18 to 35.
For each subject, the PPG signal was recorded from wrist using a  pulse oximeter   with green LED (wavelength: 515nm). The acceleration signal was also recorded from  wrist using a three-axis accelerometer. Both the pulse oximeter and the accelerometer were embedded in a wristband, which was comfortably worn. The ECG signal was recorded from the chest using wet ECG sensors. All signals were sampled at 125 Hz.

During data recording subjects walked or ran on a treadmill with the following speeds in order: the speed of 1-2 km/hour for 0.5 minute, the speed of 6-8 km/hour for 1 minute, the speed of 12-15 km/hour for 1 minute, the speed of 6-8 km/hour for 1 minutes, the speed of 12-15 km/hour for 1 minute, and the speed of 1-2 km/hour for 0.5 minute. The subjects were asked to purposely use the hand with the wristband to pull clothes, wipe sweat on forehead, and push buttons on the treadmill, in addition to freely swing.

\subsection{Parameter Settings}

When using FOCUSS for SSR, we chose the Regularized M-FOCUSS algorithm \cite{cotter2005sparse} \footnote{The algorithm was designed for an extended SSR model, but can also be used for the basic SSR model, since the latter is a special case of the former.}. We set its parameter $p=0.8$, the regularization parameter $\lambda = 0.1$, and the spectrum grid parameter $N = 4096$ (given in (\ref{eq:basis})). Note that the algorithm does not need to iterate till convergence. It has been observed that coefficients with large values can converge after only a few iterations \cite{Gorodnitsky1997}. Thus we set the iteration number to 5.

\subsection{Performance Measurement}
\label{subsec:performance}

The ground-truth of HR in each time window was calculated from the simultaneously recorded ECG signal. Given a time window, we counted the number of cardiac cycles $H$ and the duration $D$ (in seconds), and then calculated the HR as $60 H/D$ (in BPM). We did not use any ECG heart rate estimation algorithms to avoid  estimation errors.

To evaluate TROIKA's performance, multiple measurement indexes were used.
Denote by $\mathrm{BPM}_\mathrm{true}(i)$ the ground-truth of HR in the $i$-th time window, and denote by $\mathrm{BPM}_\mathrm{est}(i)$ the estimated HR using TROIKA. The average absolute  error is defined as
\begin{eqnarray}
\mathrm{Error1} = \frac{1}{W} \sum_{i=1}^W \big|\mathrm{BPM}_\mathrm{est}(i) -  \mathrm{BPM}_\mathrm{true}(i) \big|
\end{eqnarray}
where $W$ is the total number of time windows.  Similarly, the average absolute error percentage, defined as
\begin{eqnarray}
\mathrm{Error2} = \frac{1}{W} \sum_{i=1}^W \frac{\big|\mathrm{BPM}_\mathrm{est}(i) -  \mathrm{BPM}_\mathrm{true}(i) \big|}{\mathrm{BPM}_\mathrm{true}(i) },
\end{eqnarray}
was calculated. The Bland-Altman plot \cite{martin1986statistical} was also used to examine the agreement between ground-truth and estimates, which shows the difference between  each estimate and the associated ground-truth against their average. Limit of Agreement (LOA) in this analysis was calculated, which is defined as $[\mu-1.96\sigma,\mu+1.96\sigma]$, where $\mu$ is the average difference and $\sigma$ is the standard deviation. In this range 95\% of all differences are inside. Pearson correlation between  ground-truth and  estimates was also evaluated.

\subsection{Results}

\begin{figure}[t]
\centering
\includegraphics[width=8cm,height=4.5cm]{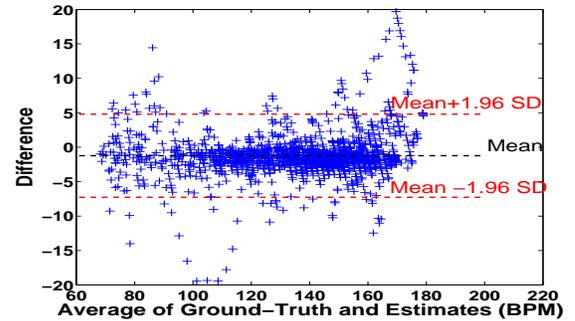}
\caption{The Bland-Altman plot of the estimation results on the 12 datasets.}
\label{fig:BAplot}
\end{figure}

\begin{figure}[t]
\centering
\includegraphics[width=8cm,height=4.5cm]{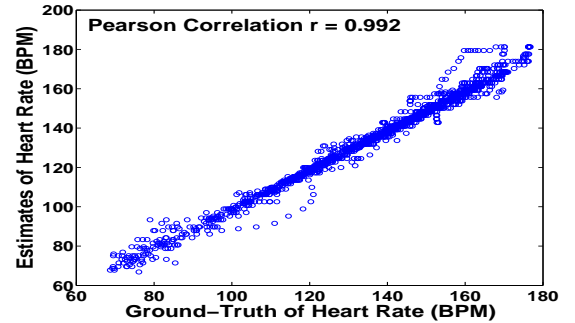}
\caption{The Pearson correlation of the estimation results on the 12 datasets.}
\label{fig:corrPlot}
\end{figure}

Table~\ref{table:err1} and Table~\ref{table:err2} list the average absolute error (Error1) and the average error percentage (Error2) on all 12 subjects' recordings, respectively. Using the complete TROIKA framework (namely all signal decomposition, sparse signal reconstruction, and spectral peak verification are employed), we obtained satisfactory results. Averaged across the 12 subjects, the absolute estimation error (Error1) was $2.34 \pm 0.82$ BPM (mean $\pm$ standard deviation), and the error percentage (Error2) was $1.80\%$. Figure~\ref{fig:BAplot} gives the Bland-Altman plot. The LOA was $[-7.26, 4.79]$ BPM (standard deviation $\sigma=3.07$ BPM). Figure~\ref{fig:corrPlot} gives the Pearson coefficient plot; the Pearson correlation coefficient was 0.992.

To highlight importance of the three main parts of TROIKA,  we considered three cases, i.e. removing SSA, replacing the SSR algorithm with FFT, and removing the spectral peak verification. We evaluated the performance in the three cases, and listed the estimation errors in Table~\ref{table:err1} and Table~\ref{table:err2} as well. The results show that if any of the three main parts was missing, the performance was not robust; TROIKA did completely lose the tracking of HR on some datasets.

\begin{figure}[t]
\centering
\includegraphics[width=9cm,height=6cm]{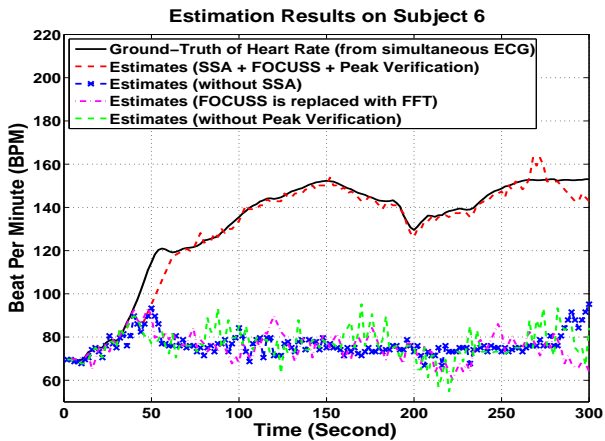}
\caption{Estimation results on recordings of Subject 6. Four cases were considered, i.e. using the complete TROIKA framework, removing the SSA part, replacing SSR with Periodogram (computed via FFT), and removing the peak tracking verification. In each case the estimated HR trace was plotted, and was compared to the ground-truth (obtained from simultaneous ECG).}
\label{fig:sub6}
\end{figure}

\begin{figure}[t]
\centering
\includegraphics[width=9cm,height=6cm]{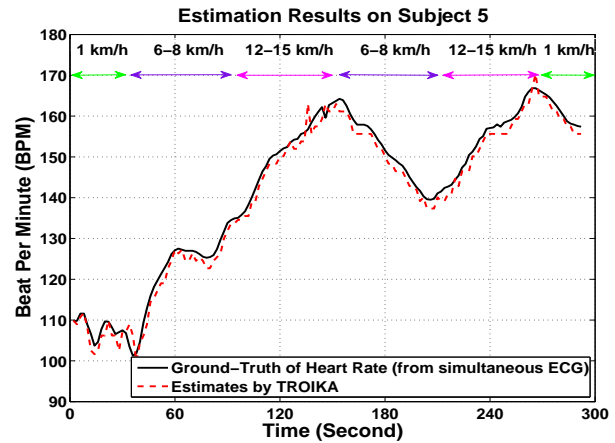}
\caption{Estimation result on recordings of Subject 5. The walking/running speed at every duration is indicated on the top of the figure.  }
\label{fig:sub5}
\end{figure}

To get closer look to the performance of TROIKA, Figure~\ref{fig:sub6} gives the result on the recordings of Subject 6, which shows that missing any part of TROIKA resulted in failure. Figure~\ref{fig:sub5} shows the result on recordings of Subject 5 as another example when all parts of TROIKA played roles. The estimated HR was very close to the ground-truth, and every small change in the ground-truth was estimated with high fidelity.

\begin{figure}[t]
\centering
\includegraphics[width=9.5cm,height=8cm]{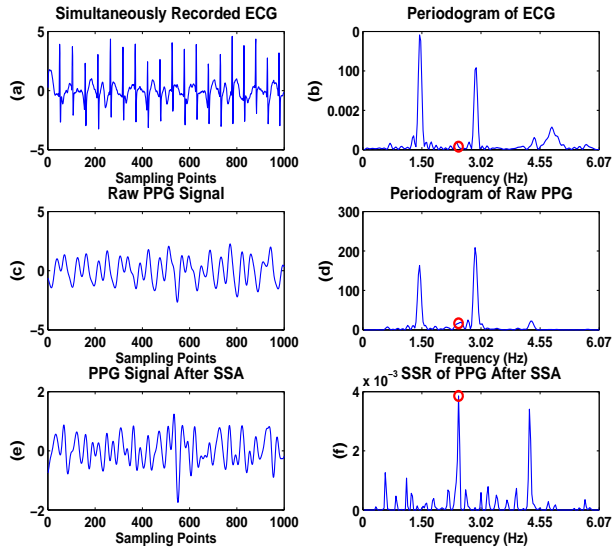}
\caption{An example showing the benefit of SSA. (a) A segment of simultaneously recorded ECG (as a ground-truth for HR estimation). (b) Spectrum (by Periodogram) of the ECG segment. (c) and (d) is a segment of simultaneously recorded raw PPG, and its spectrum (by Periodogram), respectively. (e) and (f) is the PPG after SSA, and its sparse spectrum calculated by FOCUSS, respectively. All signals in (a), (c), and (e) are normalized to have zero mean and unit variance. The circles in (b), (d) and (f) indicate the spectral peaks corresponding to the HR (which was  calculated as described in Section \ref{subsec:performance}). }
\label{fig:stepResult}
\end{figure}

To better see the benefit of using SSA to partially remove motion artifacts,  we carried out an experiment using a raw PPG segment and a simultaneous ECG segment (see Figure~\ref{fig:stepResult} (a) and (c), respectively). Both the simultaneous ECG segment and the PPG segment contained strong motion artifacts, which can be seen from their spectra (calculated by Periodogram) in Figure~\ref{fig:stepResult} (b) and (d), respectively. In their spectra, the two significant peaks correspond to the fundamental and harmonic frequencies of strides. However, the spectral peak associated with HR is not discernible. In contrast, after SSA, the spectrum of the cleansed PPG signal (Figure~\ref{fig:stepResult} (e)) clearly presents the spectral peak associated with the HR, as shown in Figure~\ref{fig:stepResult} (f). This result shows that SSA can remove motion artifacts in PPG and make more significant the spectral peak associated with HR.

\begin{figure}[t]
\centering
\includegraphics[width=8cm,height=6cm]{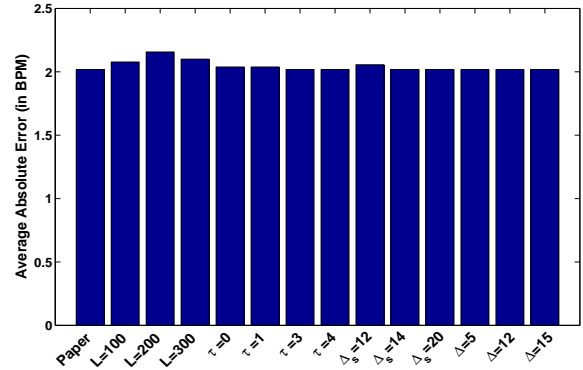}
\caption{Robustness of TROIKA to parameter values. 'Paper' indicates the result with parameter values given in Section \ref{sec:TROIKA}. 'L=100' indicates the result with all parameter values given in Section \ref{sec:TROIKA} except to $L$ which was changed to 100. Similar meanings go to other x-labels in the figure.}
\label{fig:sensitivity}
\end{figure}

To verify TROIKA is robust to parameter values, we used the recordings of Subject 5 to study the sensitivity of TROIKA to the parameters $L$ (used in SSA), $\Delta$ (used in SSA), $\tau$ (used in spectral peak tracking) and $\Delta_s$ (used in spectral peak tracking). In previous experiments these parameters were set as $L=400$, $\Delta=10$, $\tau=2$, and $\Delta_s=16$. In this experiment, TROIKA was performed using these parameter values except to one parameter which was changed to another value. In particular, $L$ was changed to 100, 200, and 300, $\Delta$ was changed to 5, 12, and 15, $\tau$ was changed to 0, 1, 3, and 4, and $\Delta_s$ was changed to 12, 14, and 20. The results, given in Figure \ref{fig:sensitivity}, show that the performance was almost the same regardless of the specific parameter values.

\subsection{Discussions}

Comparing to prior works on HR monitoring during running, TROIKA showed competitive or superior performance. For example, in \cite{poh2010motion} the Pearson correlation coefficient was 0.75 when subjects were running at peak speed of 8 km/hour. In \cite{yousefi2013ANC} the Pearson correlation coefficient was 0.78 at the speed of 3 miles/hour (4.8 km/hour), and 0.64 at the speed of 5 miles/hour (8 km/hour). In contrast, in our experimental results the correlation coefficient was 0.992. In \cite{fukushima2012SpectrumSubtraction} the standard deviation of error was 8.7 BPM, while in our experimental results the standard deviation was 3.07 BPM.

One can adopt a lower sampling rate to reduce computational load. We found when the sampling rate was even reduced to 25 Hz and some parameters were tuned suitably, the HR estimation performance was  almost the same but the running time was dramatically shortened. We will investigate this topic in future works.

\section{Conclusion}

In this paper we proposed a general framework, called TROIKA, for heart rate estimation using wrist-type PPG signals when subjects are performing intensive physical exercise. The framework consists of three key parts: signal decomposition for denoising, sparse signal reconstruction for high-resolution spectrum estimation, and spectral peak tracking with verification mechanisms. Experimental results on recordings from 12 subjects showed that TROIKA has high estimation accuracy, and each of its three parts is indispensable to the high performance.

The TROIKA framework provides large flexibility. Many variants can be derived from it by choosing specific algorithms in the three key parts according to specific requirements in hardware design. Thus it is of great values to any wearable devices.

\bibliographystyle{IEEEtran}

\bibliography{bib_PPG,bib_CS,bib_Spectrum,bib_SBL,bib_Zhilin}

\end{document}